\documentclass{osa-article}

\journal{oe}

\usepackage{amsmath}

\begin{document}

\title{Photonic Computing to Accelerate Data Processing in Wireless Communications}

\author{Mahsa Salmani\authormark{1}, Armaghan Eshaghi\authormark{2}, Enxiao Luan, Sreenil Saha}
\address{Huawei Technologies Canada, ON, L3R 5A4, Canada\\
\authormark{1}mahsa.salmani@huawei.com\\
\authormark{2}armaghan.eshaghi@huawei.com}




\begin{abstract}
Massive multiple-input multiple-output (MIMO) systems are considered as one of the leading technologies employed in the next generations of wireless communication networks (5G), which promise to provide higher spectral efficiency, lower latency, and more reliability. Due to the massive number of devices served by the base stations (BS) equipped with large antenna arrays, massive-MIMO systems need to perform high-dimensional signal processing in a considerably short amount of time. The computational complexity of such data processing, while satisfying the energy and latency requirements, is beyond the capabilities of the conventional widely-used digital electronics-based computing, i.e., Field-Programmable Gate Arrays (FPGAs) and Application-Specific Integrated Circuits (ASICs). In this paper, the speed and lossless propagation of light is exploited to introduce a photonic computing approach that addresses the high computational complexity required by massive-MIMO systems. The proposed computing approach is based on photonic implementation of multiply and accumulate (MAC) operation achieved by broadcast-and-weight (B$\&$W) architecture. The B$\&$W protocol is limited to real and positive values to perform MAC operations. In this work, preprocessing steps are developed to enable the proposed photonic computing architecture to accept any arbitrary values as the input. This is a requirement for wireless communication systems that typically deal with complex values. 
Numerical analysis shows that the performance of the wireless communication system is not degraded by the proposed photonic computing architecture, while it provides significant improvements in time and energy efficiency for massive-MIMO systems as compared to the most powerful Graphics Processing Units (GPUs).
\end{abstract}

\section{Introduction}
\label{intro}
The next generations of wireless communication networks, i.e., 5G and Beyond (5GB), are designed to accommodate a large number of smart devices, connected to each other, that are being served by base stations equipped with a massive number of antennas while the requirements of individual devices on the reliability, latency, and energy consumption are met \cite{gupta2015survey, saad2019vision}. Massive-MIMO systems are one of the key enablers of 5GB to provide high-data-rate and low-latency connectivity \cite{navarro2020survey}. 
Such dense connectivity together with the demands for higher data rate and lower latency will increase the complexity, and accordingly, the cost of data processing in 5GB systems. In massive-MIMO systems, due to the large number of antennas at the BS and massive number of mobile devices in the network, parallel signal processing for missions such as channel estimation, precoding, and signal detection will become increasingly complex and time-consuming. This complexity has been considered as of the main bottlenecks of realizing massive-MIMO systems. Accordingly, efficient methods for reducing this complexity and improving the efficiency of the radio transceiver architectures are required. Different optimization techniques have been proposed in the literature to address the 5GB challenges from both algorithmic and hardware implementation perspectives \cite{doan2004design, prabhu2015algorithm, sohrabi2016hybrid, heath2016overview, li2017decentralized}. Moreover, machine learning techniques have been recently exploited to reduce the ever-increasing computational complexity and to satisfy ultra-reliability, high data rate, and low latency requirements of 5GB \cite{sun2019application, chen2019artificial}. It has also been shown that a careful co-design of algorithms and hardware parameters can result in an even more energy-efficient signal processing in MIMO systems \cite{van2018efficient,zhang2016large}.

While the proposed methods can result in a less complex signal processing for massive-MIMO systems, the limitations imposed by the digital electronics-based hardware preclude full exploitation of the improvements offered by those algorithms. In particular, in order to provide the desired connectivity in 5GB massive-MIMO systems, the base stations are potentially equipped with more than thousands of antennas to serve more than hundreds of mobile users in the network. In these scenarios, precoding the signals that are to be transmitted to the users or detecting the signals received from the users at the BS requires billions of MAC operations per second. For millimeter-wave (mmWave) 5G, in which the latency requirements limit the slot length to be as short as tens of millisecond, the required rate for MAC operations to perform tasks such as precoding or detection is in the order of hundreds of Tera MACs per second. This number can easily reach to tens of Peta MACs per second for beyond-5G wireless communications. In addition, due to the mobility of the users and the dynamics of the environment, the channel state information matrix, and accordingly, the precoding and/or detection matrix need to be updated frequently \cite{Prabhu2014}. Such computational requirements can be hardly met with current power- and bandwidth-limited base stations with digital electronics-based processing units. Computing based on digital electronics hardware which is accompanied by components such as analog-to-digital (ADC) and digital-to-analog (DAC) converters faces fundamental challenges in terms of processing rate and energy consumption \cite{sundstrom2008power, jebashini2015survey}. On the other hand, the fact that analog electronic components are frequency-dependent with poor reconfigurability in the radio frequencies (RF) limits their application in 5GB systems \cite{mongia2007rf}.

Photonics-based computing has been proposed as a promising approach to provide high-performance and low-latency systems for large-scale signal processing \cite{5645652, capmany2012microwave, vlasov2012silicon}. In particular, an integrated optical platform comprising of both active elements such as modulators, lasers and photodetectors, and passive elements such as waveguides and couplers has shown orders of magnitude improvement in computation time and throughput as compared to the electronics counterparts \cite{de2017progress}. Leveraging unique features of light, optical computing has been regarded as one of the emerging technologies to address the ``von-Neumann bottleneck'' \cite{miscuglio2020photonic}.

One of the recently-developed photonics-based computing protocols is Broadcast-and-Weight (B$\&$W) \cite{tait2014broadcast}, in which wavelength-division multiplexing (WDM) scheme, a bank of microring modulators (MRM), and balanced photodetectors (PD) are utilized to implement weighted addition in a photonic platform.
This photonic MAC unit can provide significant potential improvements over digital electronics in energy, processing speed, and compute density \cite{de2017progress}. However, the proposed architecture \cite{tait2014broadcast} has limitations that makes it unqualified for communication systems. One of the most important limitations is that the B$\&$W architecture can only realize real-valued vectors or matrices. Moreover, it is not capable of operating MAC over two negative-valued inputs. Finally, there exist constraints on the number of MRMs and parallel wavelength channels that can be realized in the system. Therefore, the typical large matrices in communication systems need to be partitioned in an efficient way before processing. 

In this paper, we exploit the B$\&$W architecture to develop a photonic computing platform that meets the stringent requirements of next-generation wireless communication systems, including massive-MIMO-enabled networks. The proposed photonic computing platform tackles the aforementioned limitations of the B$\&$W architecture, and hence, it is capable of supporting wireless communication networks. In particular, we devise simple preprocessing steps by which inputs (vectors or matrices) with arbitrary values (real or imaginary, positive or negative) can fit into the proposed computing platform. Furthermore, by utilizing different algorithmic approaches such as matrix inversion approximation and parallelization techniques, the efficiency of the proposed architecture for matrix inversion and large-size matrix multiplication is improved. Several numerical analyses show that while the performance of the proposed photonic architecture is comparable to the performance of the digital-electronics processing units such as GPUs, the time efficiency of the proposed architecture is significantly improved.   

\section{Massive-MIMO Systems}
\subsection{System Model}
\label{system_model} 
The computational complexity required for signal detection in a massive-MIMO system is formulated in this section and the capability of digital electronics and optics to address such computational requirement is explored accordingly. Consider a massive-MIMO uplink system with $K$ single-antenna users and a BS equipped with $M$ antennas, where the channels between the users and the BS are modelled as block Rayleigh fading channels. If $\mathbf{x} \in \mathbb{C}^{K\times1}$ denotes the vector of the transmitted symbols, that are selected from finite modulation set $\mathcal{S}$, i.e., $\mathbf{x} =\{ x_i | x_i \in \mathcal{S}\}$, and $\mathbf{n}$ denotes the white symmetric Gaussian noise, i.e., $n \sim \mathcal{CN}(0, \sigma^2)$, then the received signal, $\mathbf{y} \in \mathbb{C}^{M\times1}$ , can be written as 
\begin{equation}
\mathbf{y}=  \mathbf{H} \mathbf{x} +\mathbf{n},
\end{equation}
where matrix $\mathbf{H} \in\mathbb{C}^{M\times K}$ denotes the channels between the users and the BS, which is assumed to be perfectly known at the transmitter and the receiver.

\subsection{Signal Detection in MIMO Systems}
After the signals transmitted by the users are received, the BS is to obtain the best estimation of the signal transmitted by each user, $\hat{\mathbf{x}}$, by solving the following optimization problem,
\begin{equation}
\label{ML_solution}
\hat{\mathbf{x}} = \arg \min_{\mathbf{x}} \| \mathbf{y}-\mathbf{H} \mathbf{x} \|.
\end{equation}
The optimal solution of the optimization problem in \eqref{ML_solution} can be obtained by the Maximum Likelihood (ML) detection. However, since the computational complexity of the ML detection increases exponentially by increasing the number of antennas, alternative  detection  schemes, such as Zero Forcing (ZF) or Minimum Mean-Square Error (MMSE) have been proposed in the literature \cite{rusek2012scaling}. Depending on the detection scheme that is employed by the massive-MIMO system, a detection matrix, namely $\mathbf{A}$, is constructed and used to detect the transmitted signal through linear processing as follows
\begin{equation}
\label{linear_detection}
\hat{\mathbf{x}} = \mathbf{A}\mathbf{y} = \mathbf{A}\mathbf{H} \mathbf{x}+ \mathbf{A}\mathbf{n},
\end{equation}
where
\begin{equation}
\label{lin_detections}
    \begin{cases}
      \mathbf{A} = (\mathbf{H}^{H}\mathbf{H})^{-1}\mathbf{H}^{H}, & \text{in ZF},   \\
      \mathbf{A} =(\mathbf{H}^{H}\mathbf{H} +\sigma^2\mathbf{I})^{-1}\mathbf{H}^{H}, & \text{in MMSE}.
    \end{cases}       
\end{equation}

\subsection{Computational Complexity of Signal Detection in Massive-MIMO Systems}
\label{sec_complexity}
The complexity of signal processing in 5GB can be mostly attributed to computationally-complex matrix operations such as multiplications and inversions of matrices with large sizes (see \eqref{lin_detections}). In order to measure the complexity of signal processing required in massive-MIMO systems, the number of Floating Point operations (FLOPs) \cite{hunger2005floating} or the number of MAC operations can be calculated (note that each MAC operation includes two FLOPs). 
As can be seen in \eqref{lin_detections}, the linear detection process involves matrix inversions and matrix multiplications which both have complexity of the order $O(K^3)$ MACs (FLOPs). Accordingly, the computational complexity in a wireless communication system scales with the number of antennas at the BS, or with the number of users in the cellular network, or both. This is specifically a fundamental challenge for 5GB, where massive-MIMO is an essential part of the development. 

\subsubsection{Digital and Analog Electronics}
Consider a massive-MIMO system in which the BS with more than a thousand antennas is serving more than a hundred mobile users. In this system, according to \eqref{lin_detections}, the signal detection requires about one million MAC operations, while in order to meet the latency requirement the slot length should be as low as 125 $\mu$s. Thus, the required MAC operations rate to complete the detection task is more than 440 TMACs/s. The BS must be able to process the data at such rate while keeping the power consumption and the size of the data processing unit equitable.  

In recent years, GPUs and FPGAs have been developed to encompass general purpose tasks in the high-performance computing arena. The most powerful GPU architecture to date is \textit{NVIDIA VOLTA}\footnote{\url{https://www.nvidia.com/en-sg/data-center/volta-gpu-architecture/}} with 640 tensor cores and 21 billion transistors which can deliver more than 50 TMACs/S. However, tens of these units are needed to meet the required computational power. In these systems, I/O latency and sequential processing capabilities cannot exceed the time resolution of the processor which is ultimately bounded by its clock rate.

To tackle the speed limitation of digital-electronics-based processors while maintaining a reasonable area and power consumption, an optical computing approach is proposed in this paper as a revolutionary computing paradigm. It allows very complex operations to be performed in real time, which can significantly offload electronic post-processing and provide a technology to make RF decisions on-the-fly.  

\section{Photonic Computation for Massive-MIMO Systems}
\begin{figure}[htbp]
\centering \includegraphics[width=10cm]{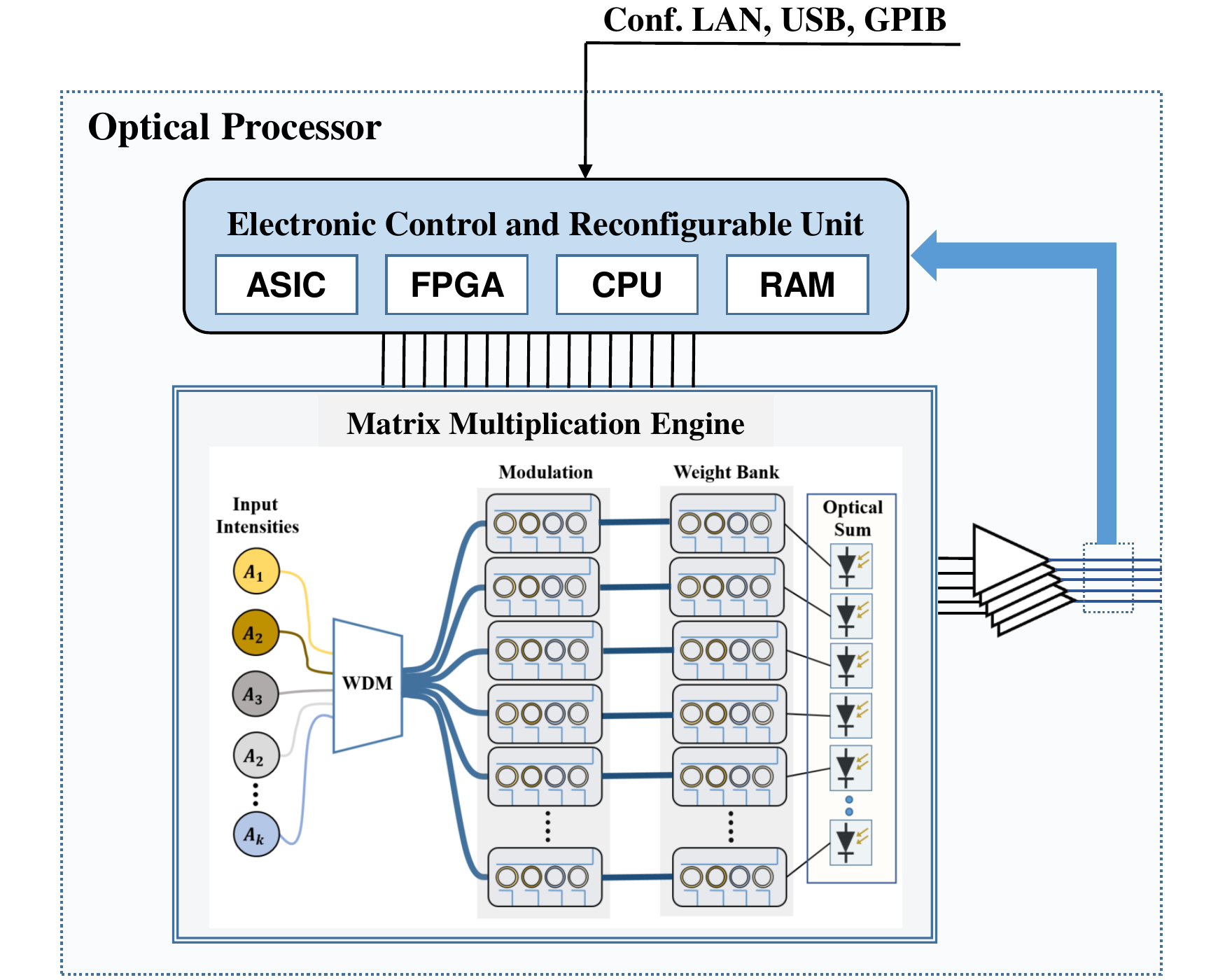}
 \caption{Schematic of the proposed photonic computing platform.}
\label{Generic_architecture}
\end{figure}

The architecture of the proposed photonic computing platform for ultra-fast signal processing in the next generations of wireless communication networks is depicted in Figure~\ref{Generic_architecture}. The processor core which is based on the B$\&$W architecture \cite{tait2014broadcast} is a photonic-integrated circuit (PIC) fabricated on a silicon photonics (SiPh) platform. It contains a matrix-multiplication engine where the input vectors are loaded using microring resonators (MRRs) in the modulation and weight bank sections. The photodetector array performs the optical summation. Wavelength-division multiplexing scheme is adopted where all optical inputs, spatially separated by wavelength, lies in a single waveguide. The Electronic Control and Reconfigurable Unit (ECRU) is composed of interconnected ASIC, FPGA, central processing unit (CPU) and random-access memory (RAM) modules. Its main function is to generate analog control signals for setting the weights of the microring photonic modulators. Another important feature of the ECRU unit is to make sure that the processor core is well calibrated by correcting for the fabrication variations and regulating the controls signals against any thermal fluctuations. By means of the General-Purpose Input/Output (GPIO) or Universal Serial Bus (USB), the ECRU unit maintains a high-bandwidth communication link with a computer motherboard.

\subsection {Matrix-Multiplication Engine}
\label{Sec:mat_to_mat}
\begin{figure}[htbp]
\centering
\includegraphics[width=10cm]{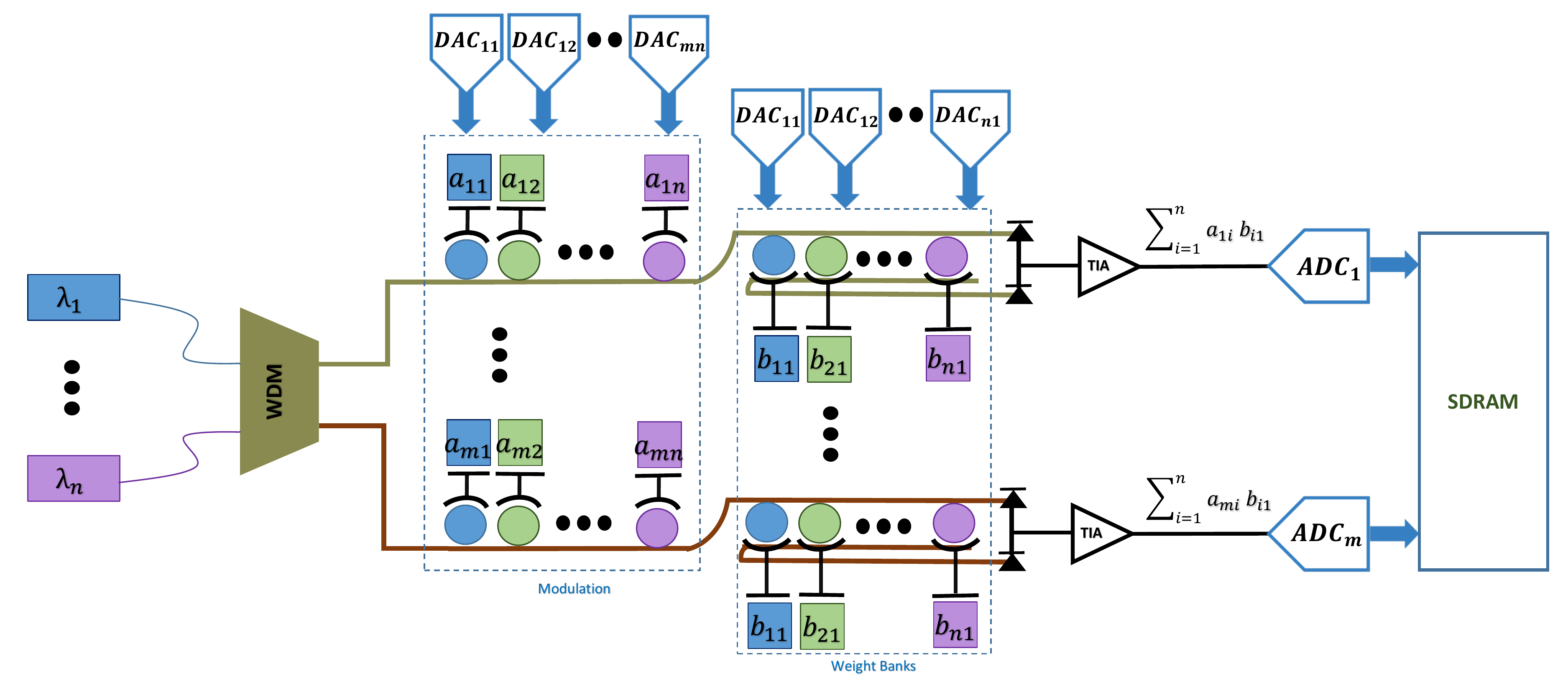}
 \caption{Implementing matrix-to-vector multiplication with B$\&$W architecture.}
\label{multiply_mat_to_mat}
\end{figure}

As depicted in Figure~\ref{multiply_mat_to_mat}, in order to implement matrix multiplication, i.e. $\mathbf{A} \times \mathbf{B}$, where $\mathbf{A} \in \mathbb{R}_{+}^{m\times n}$ and $\mathbf{B} \in \mathbb{R}^{n\times k}$, in the proposed optical architecture, the elements of the first vector-to-be-multiplied are loaded using all-pass MRMs and are encoded in the intensities of the wavelength-multiplexed signals (Modulation section). The elements of the second vector-to-be-multiplied are encoded as weights using add-drop MRMs (Weight Bank section). 
The interfacing of optical components with electronics are facilitated by the use of mixed-signal integrated circuit blocks such as DACs and ADCs, integrated inside the ASIC in the ECRU unit. The multiplication is performed by linking the elements of the first and the second vectors via an optical waveguide, and the accumulation is performed by the photodetector followed by a transimpedance amplifier (TIA) to provide electronic gain, which is also integrated in the same ASIC. For heterogeneous integration, the different analog and digital electronic control circuitry such as ADCs, DACs, TIAs are fabricated in a standard complementary metal–oxide–semiconductor (CMOS) process and interfaced with the corresponding SiPh chip by means of wire-bonding or flip-chip bonding. The MRMs are controlled by the DACs, while the interfacing with ADC is required to compute the digital representation of the analog output, which can then be stored in the SDRAM and processed by the CPU or FPGA.

In our numerical simulations, the number of wavelength channels is considered to be equal to the number of rows of the left-hand-side (LHS) matrix, $m$, and we set the number of MRRs in the Modulation and Weight Bank sections, equal to the number of columns of the LHS matrix or number of rows of the right-hand-side (RHS) matrix, $n$. 

\subsection{Photonic Matrix Inversion}
In addition to matrix multiplication, matrix inversion is a widely-used operation in wireless communication networks. In 5GB, the inverse of an arbitrary, potentially large, and complex matrix needs to be calculated on-the-fly. Although the computational complexity of matrix multiplication and matrix inversion are in the same order, matrix multiplication is preferred from the hardware implementation perspective \cite{Prabhu2014}. Accordingly, several algorithms have been proposed in the literature to approximate the inverse of a matrix with a set of matrix multiplications, e.g., conjugate gradients \cite{wu2016} and iterative algorithms such as Newton \cite{tang2016} and Neumann series method \cite{Prabhu2013, fang2016}. Cholesky factorization can also be considered as another technique that can be used to implement matrix inversion with a number of matrix multiplications \cite{Perre2018}. Iterative algorithms can outperform other approaches by taking advantage of reusing available resources in the processing unit, and hence are considered to be more hardware-friendly \cite{thanos2017algorithms}. 
 
In massive-MIMO systems calculating the inverse of the Gram matrix, $\mathbf{H}^H \mathbf{H}$, is an essential operation. In a massive-MIMO system with $K$ users and $M$ antennas at the BS, where $M\gg K$, the Gram matrix becomes diagonally dominant, and that leads to an accurate approximation of inverse matrix in both Newton and Neumann series approximations.  

In the following, calculating the inverse of matrix $\mathbf{A} \in \mathbb{C}^{K\times K}$ using Newton iterative method and Neumann-series approximation techniques is explained.

\subsubsection {Neumann-Series Approximation}
\label{Neumann}
The Neumann-series expansion of the inverse of a matrix $\mathbf{A}$ is given as \cite{Prabhu2013, fang2016}
\begin{equation}
\label{neumann_eq}
\hat{\mathbf{A}}^{-1} = \sum_{n=0}^{\infty} (\mathbf{X}^{-1} (\mathbf{X}-\mathbf{A}))^n \mathbf{X}^{-1},
\end{equation}
where $\hat{\mathbf{A}}^{-1}$ is guaranteed to converge to the exact inverse of matrix $\mathbf{A}$ when 
\begin{equation}
\label{neumann_const}
\lim_{n \rightarrow \infty} (\mathbf{I}- \mathbf{X}^{-1} \mathbf{A})^n = \mathbf{0}.
\end{equation}
In massive-MIMO systems, where $\mathbf{A}=\mathbf{H}^H \mathbf{H}$ is a diagonally-dominant matrix, the condition in \eqref{neumann_const} holds. In that case, if matrix $\mathbf{A}$ is rewritten as
\begin{equation}
\label{diag_A}
\mathbf{A} = \mathbf{A}_{\text{diag}} + \mathbf{A}_{\text{off-diag}},
\end{equation}
where matrix $\mathbf{A}_{\text{diag}}$ is a diagonal matrix with diagonal elements of $\mathbf{A}$ and matrix $ \mathbf{A}_{\text{off-diag}}$ holds all elements of the matrix $\mathbf{A}$ and has zeros on the main diagonal, the $K$-term Neumann series approximation is
\begin{equation}
\label{neumann_eq}
\hat{\mathbf{A}}^{-1}_{K} = \sum_{n=0}^{K}  \bigl(- \mathbf{A}_{\text{diag}}^{-1} \mathbf{A}_{\text{off-diag}} \bigr)^n \mathbf{A}_{\text{diag}}^{-1}.
\end{equation}

\subsubsection{Newton Approximation Method}
\label{Newton}
For an arbitrary invertible matrix $\mathbf{A}$, with an initial rough estimation of its inverse $\mathbf{X}^{-1}_0$, the estimated inverse matrix at the $n^{\text{th}}$ iteration of Newton approximation technique is \cite{tang2016}
\begin{equation}
\label{newton_eq}
\mathbf{X}^{-1}_n = \mathbf{X}^{-1}_{n-1} (2 \mathbf{I}-\mathbf{A} \mathbf{X}^{-1}_{n-1}),
\end{equation}
where  $ \mathbf{A}_{\text{diag}}^{-1}$, which is a diagonal matrix with diagonal elements of $\mathbf{A}$, can be used as the first rough estimation, $\mathbf{X}^{-1}_0$.
The main advantage of the Newton method is that it converges quadratically to the inverse matrix if $\| \mathbf{I}- \mathbf{A}\mathbf{X}^{-1}_0\| < 1$.
This condition is satisfied for the diagonally-dominant Gram matrix in the uplink data detection in massive-MIMO systems.

\subsection{Algorithm-Hardware Co-Design for Photonic Computing}
\label{Preprocessings}
In B$\&$W architecture, the elements of the LHS matrix are encoded into the light intensities. This implies that only real positive values can be realized in the this architecture and it limits the application of the B$\&$W-based photonic MAC in a variety of cases including wireless communication networks. In order to tackle this issue, in the following sections, preprocessing steps are proposed such that any arbitrary matrix can be represented by the optical architecture.

\subsubsection{Preprocessing Step 1: Addressing Complex-valued Matrices}
In order to represent complex-valued matrices with the proposed photonic computing platform, the real representation of complex-valued matrices is explored. Any arbitrary complex-valued matrix, $\mathbf{A} \in \mathbb{C}^{m \times n}$, can be written as the summation of the real and imaginary parts,
\begin{equation}
\label{real_imag_mat}
\mathbf{A} = \mathbf{A}_{r}+ j\mathbf{A}_{i},
\end{equation}
where $\mathbf{A}_{r} \in \mathbb{R}^{m \times n}$ and $\mathbf{A}_{i} \in \mathbb{R}^{m \times n}$ are real-valued matrices denoting the real and imaginary parts of $\mathbf{A}$, respectively. 
According to \eqref{real_imag_mat}, the multiplication of two complex-valued matrices, namely $\mathbf{A} \in \mathbb{C}^{m \times n}$ and $\mathbf{B} \in \mathbb{C}^{n \times k}$, can be obtained as
\begin{subequations}
\begin{align}
\mathbf{A} \times\mathbf{B} = & (\mathbf{A}_{r}+ j\mathbf{A}_{i}) \times (\mathbf{B}_{r}+ j\mathbf{B}_{i}) \\ 
= &  \underbrace{(\mathbf{A}_{r} \times \mathbf{B}_{r} -\mathbf{A}_{i} \times \mathbf{B}_{i} )}_\text{real part} +j \underbrace{(\mathbf{A}_{r} \times \mathbf{B}_{i} +\mathbf{A}_{i} \times \mathbf{B}_{r})}_\text{imaginary part}.
\end{align}
\end{subequations}
Therefore, in order to multiply two complex-valued matrices, four parallel real-valued matrix multiplications of the same size as that of the original matrices can be considered.

\subsubsection{Preprocessing Step 2: Addressing Negative-valued Matrices}
The proposed solution to represent negative-valued LHS matrices in this architecture is to project the negative sign of the elements of the LHS matrix, $\mathbf{A}$, to the sign of the RHS matrix, $\mathbf{B}$. In doing so, the negative-valued matrix $\mathbf{A} \in \mathbb{R}^{m \times n}$ is rewritten as a subtraction between two positive-valued matrices, $\bar{\mathbf{A}} \in \mathbb{R}^{m \times n}_{+}$ and $|a_{\text{min}}| \mathbf{1}$, where $a_{\text{min}}$ is the element of matrix $\mathbf{A}$ with the smallest value, and $\mathbf{1}$ is an all-one matrix of size $m \times n$. The negative sign of the subtraction can then be projected into the sign of the RHS matrix elements as follows 
\begin{equation}
\mathbf{A} \times\mathbf{B} = (\bar{\mathbf{A}}-|a_{\text{min}}|\mathbf{1}) \times \mathbf{B}  = \bar{\mathbf{A}}\times \mathbf{B} + |a_{\text{min}}|\mathbf{1} \times (-\mathbf{B}).
\end{equation}
Therefore, negative-valued matrix multiplication can be performed by summation of two positive-valued matrix multiplications which can be processed in parallel in the proposed photonics-based computing architecture.

\subsubsection{Preprocessing Step 3: Parallelization and Matrix Tiling Based on the Photonic Computing Architecture}
\label{sec_tiling}
The last preprocessing step is proposed to implement parallelization and matrix tiling methods. This step is required so that matrices with any arbitrary size can be processed efficiently using the proposed photonic computing system with limited number of cascaded modulators and parallel channels. 

\begin{figure}[htbp]
\centering
\includegraphics[width=0.6\textwidth]{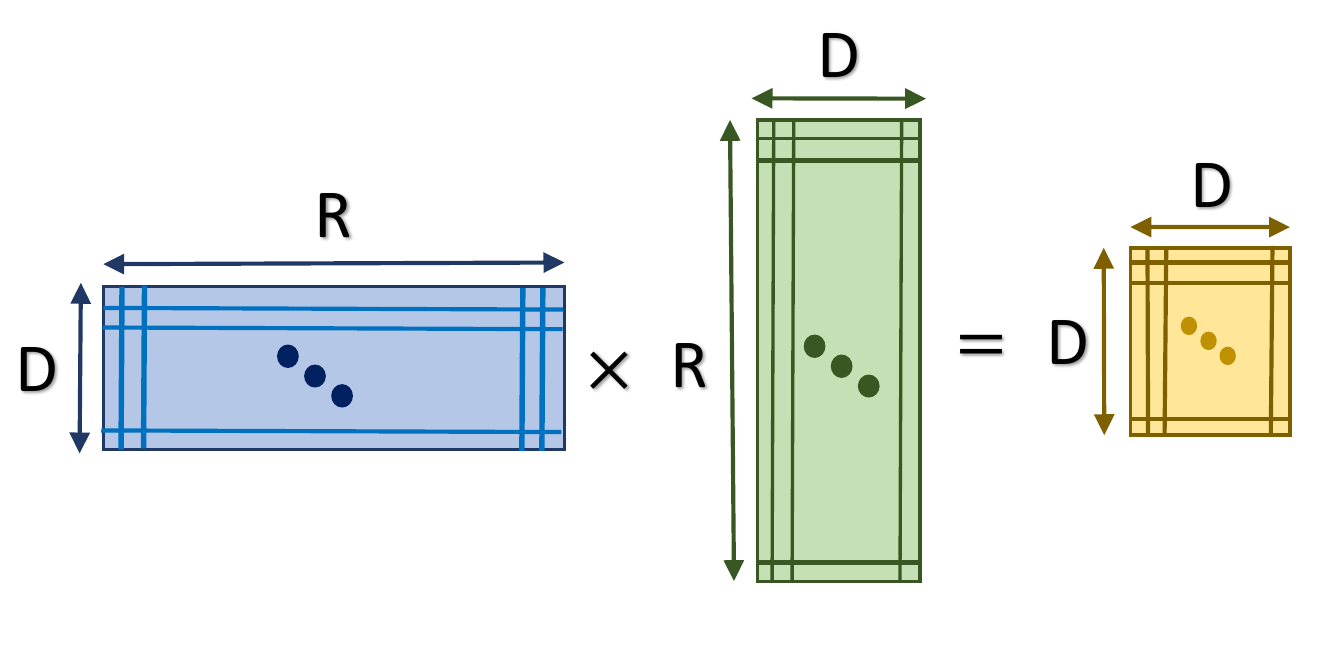}
 \caption{Size of the LHS and RHS matrices supported for multiplication using an architecture with $D$ parallel waveguide channels and $2R$ MRRs in each channel.}
\label{DR_matrix_Support}
\end{figure}

\begin{figure}[htbp]
\centering
\includegraphics[width=1\textwidth]{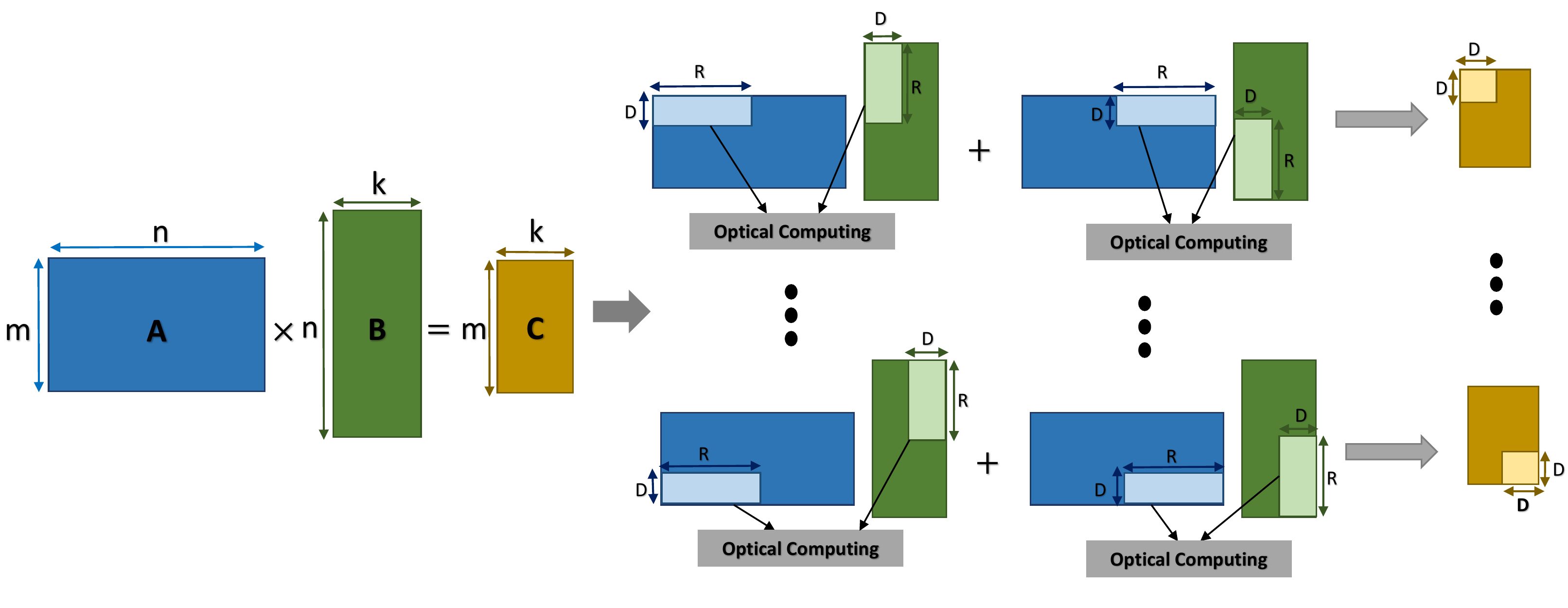}
 \caption{Matrix tiling facilitates multiplication of matrices with any arbitrary size.}
\label{matrix_tiling}
\end{figure}

Consider a computing unit with $D$ parallel wavelength channels, $R$ all-pass MRRs in the Modulation section (representing LHS matrix), and accordingly, $R$ add-drop MRRs in the Weight Bank section (representing RHS matrix) (Figure~\ref{multiply_mat_to_mat}).Each single usage of such architecture can compute multiplication of matrices with sizes that lie within the range of parameters $D$ and $R$; see Figure~\ref{DR_matrix_Support}. In order to optimally utilize this architecture to perform the multiplication between arbitrary matrices, $\mathbf{A} \in \mathbb{R}_{+}^{m \times n}$ and $\mathbf{B} \in \mathbb{R}_{+}^{n \times k}$, the matrices need to be partitioned based on $D$ and $R$, following the mapping discussed in Section~\ref{Sec:mat_to_mat}. The results of the partial multiplications are recorded in the memory and in the last step, corresponding parts are added together to generate the final result. Figure~\ref{matrix_tiling} illustrates the implementation of the multiplication of matrices $\mathbf{A}$ and $\mathbf{B}$, in which, without loss of generality, it is assumed that $m$ and $n$ are dividable into $D$ and $R$, respectively.
\section{Numerical Analysis}

The time and power efficiency of the proposed photonic computing approach mainly depend on the number of  parallel multiplexed waveguide channels, $D$, and the number of different wavelengths that can be realized in each of those channels, $R$. An optimally designed MRR is capable of supporting up to 108 WDM channels, taking both the finesse of the resonator and the channel spacing in linewidth-normalized units into account \cite{tait2019silicon}. Here, we consider a finesse of $\mathcal{F} =368$ for the MRRs, and a minimum channel spacing of 3.41 $\times$ linewidth, based on the assumption that a 3dB cross-weight penalty is allowed \cite{tait2018silicon}. Accordingly, the maximum number of the MRRs, in each of the modulation and weight-bank parts, will be $R=100$. 

An architecture with $D$ multiplexed waveguide channels contains $2RD$ MRRs in total. However, assuming that a maximum of 1024 MRRs can be manufactured in the optical architecture \cite{Bangar2020}, there is a finite set of feasible values for each of $D$ and $R$. The optimal arrangement of the number of parallel waveguide channels and the number of MRRs fundamentally depends on the computational and energy requirements. 
 
In this section, the performance and the efficiency of the proposed photonic computing platform in different practical scenarios are evaluated and compared to those of the conventional digital electronics-based processing units. 

\subsection{Power Consumption}
\label{num_power}
The total power consumption of the proposed photonic computing architecture can be obtained by adding up the power usage of different photonic and electronic components. In the architecture proposed here, $R$ lasers are utilized for generating $R$ different wavelengths, each with 100 mW power usage. The architecture contains $2DR$ MRRs and $2DR$ DACs each with 19.5 mW and 26 mW power consumption, respectively. Finally, the TIA with 17 mW and the ADC with 76 mW power usage are integrated at the output of each waveguide. Hence, the total power consumption will be calculated as

\begin{equation}
\label{power_consum}
P_{\text{Total}} (mW)  = 100 R + 91 DR + 93 D.
\end{equation}

Using \eqref{power_consum}, the power usage of the proposed architecture is calculated and reported in Table~\ref{Table_Powers}, along with those of other computation hardware baselines \cite{Bangar2020}. The results show that the power consumption of the photonic system is close to that of digital processing hardware such as GPUs. Furthermore, based on the computational requirements of the task to be executed, the power consumption of the photonic computing system can be $1/3$ of that of the best digital electronic processors in the literature. 
\begin{table}
\begin{center}
\caption{Power consumption of the proposed photonic computing system with different parameters, and the benchmarked GPUs}
    \label{Table_Powers}
 \begin{tabular}{||c c||} 
 \hline
\textbf{GPU} & \textbf{Power Usage} (W)\\ [0.5ex] 
 \hline\hline
AMD Vega FE & 375 \\ 
 \hline
AMD M125 & 300 \\
 \hline
NVIDIA Tesla V100 & 250\\
 \hline
 NVIDIA GTX 1080 Ti & 250 \\
 \hline
Photonic System ($D=32$, $R =32$) & 100 \\  
 \hline
Photonic System ($D=64$, $R =32$) & 195 \\ 
 \hline
Photonic System ($D=64$, $R =64$) & 385 \\  [1ex] 
 \hline
\end{tabular}
\end{center}
\end{table}

\subsection{Time Efficiency }
\label{sec_runtime}
The computation time of the proposed architecture mainly depends on the bandwidth of the components and the time that it takes for light to propagate through the architecture. The propagation time after multiplexing, when $2R$ MRRs are considered in each waveguide channel is estimated by 
\begin{equation}
\label{propag_time}
t_{\text{p}} = \frac{2r_{\text{MRR}}\times 2R+ 2 \times 2\pi r_{\text{MRR}}\times (\mathcal{F}/2\pi)}{(c/n_{eff})},
\end{equation}
where $r_{\text{MRR}}$ is the radius and $\mathcal{F}$ is the finesse of the MRRs. $c$ and $n_{eff}$ denote the speed of light and the effective refractive index of the waveguide, respectively. For an architecture with $2R$ cascaded MRRs, light propagates through the shared bus waveguide with the minimum length of $2r_{\text{MRR}}\times 2R$ and will be trapped by the in-resonance MRRs (only two in total, one in the modulation section and one in the weigh-bank section) $\mathcal{F}/2\pi$ times \cite{bogaerts2012silicon}. Accordingly, if $2R=200$, $r_{\text{MRR}} = 10~\mu m$, $\mathcal{F} = 368$, and $n_{eff} =2.4$, the propagation time is calculated to be $110~ps$. 
The throughput of the other components integrated in the proposed architecture is provided in Table~\ref{Table_component_speed} \cite{Bangar2020}. SDRAM is connected to a computer and is considered as digital memory before DACs and after ADCs. According to Table~\ref{Table_component_speed}, the speed is mainly limited by ADCs, DACs, and TIAs with a throughput of $10~GS/s$. Hence, the approximate computation time is equal to $100~ps$ for each single usage of the photonic system, namely, $T_{\text{single use}} = 100~ps$. 

\begin{table}
\begin{center}
\caption{The throughput and the associated processing time for different components of the proposed architecture}
    \label{Table_component_speed}
 \begin{tabular}{||c c c||} 
 \hline
\textbf{Component} & \textbf{Throughput} (GS/s) & \textbf{Processing Time} (ps)  \\ [0.5ex] 
 \hline\hline
MRR & 60 & 17 \\
 \hline
ADC &10 & 100 \\ 
 \hline
DAC & 10 & 100 \\
 \hline
Balanced PD & 25 & 40 \\
 \hline
TIA & 10 & 100 \\ 
\hline
 GDDR6 SDRAM &  16 & 60 \\ [1ex] 
 \hline
\end{tabular}
\end{center}
\end{table}

In order to obtain the total processing time for multiplication of matrices $\mathbf{A} \in \mathbb{C}^{m \times n}$ and $\mathbf{B} \in \mathbb{C}^{n \times k}$, the number of times that the architecture is used to calculate the final result needs to be calculated which depends on the dimensions of the matrices. Moreover, the upper-bound of the processing time is obtained by assuming that both the first and the second preprocessing steps in Section~\ref{Preprocessings} are required. Considering an optical chip with $D$ parallel waveguide channels, and $R$ MRRs to represent the corresponding elements of each of the matrices, the number of times that chip should be used to compute the multiplication of $\mathbf{A}$ and $\mathbf{B}$, namely $N_{\text{use}}$, can be obtained (see Figure~\ref{matrix_tiling}) as 
\begin{equation}
\label{chip_usage}
N_{\text{use}} = 8 \times k \times \lceil \tfrac{m}{D} \rceil \times \lceil \tfrac{n}{R} \rceil, 
\end{equation}
and accordingly, the total processing is calculated as
\begin{equation}
\label{total_runtime}
T_{\text{total}} (ps) = N_{\text{use}} \times T_{\text{single use}} = 8 \times k \times \lceil \tfrac{m}{D} \rceil \times \lceil \tfrac{n}{R} \rceil \times 100. 
\end{equation}

The total processing time in \eqref{total_runtime} and the power usage calculated in \eqref{power_consum} show that designing the optimal architecture requires optimization over $D$ and $R$ parameters considering the requirements of the target application.

\begin{figure}[htbp]
\centering
    \includegraphics[width=0.8\textwidth]{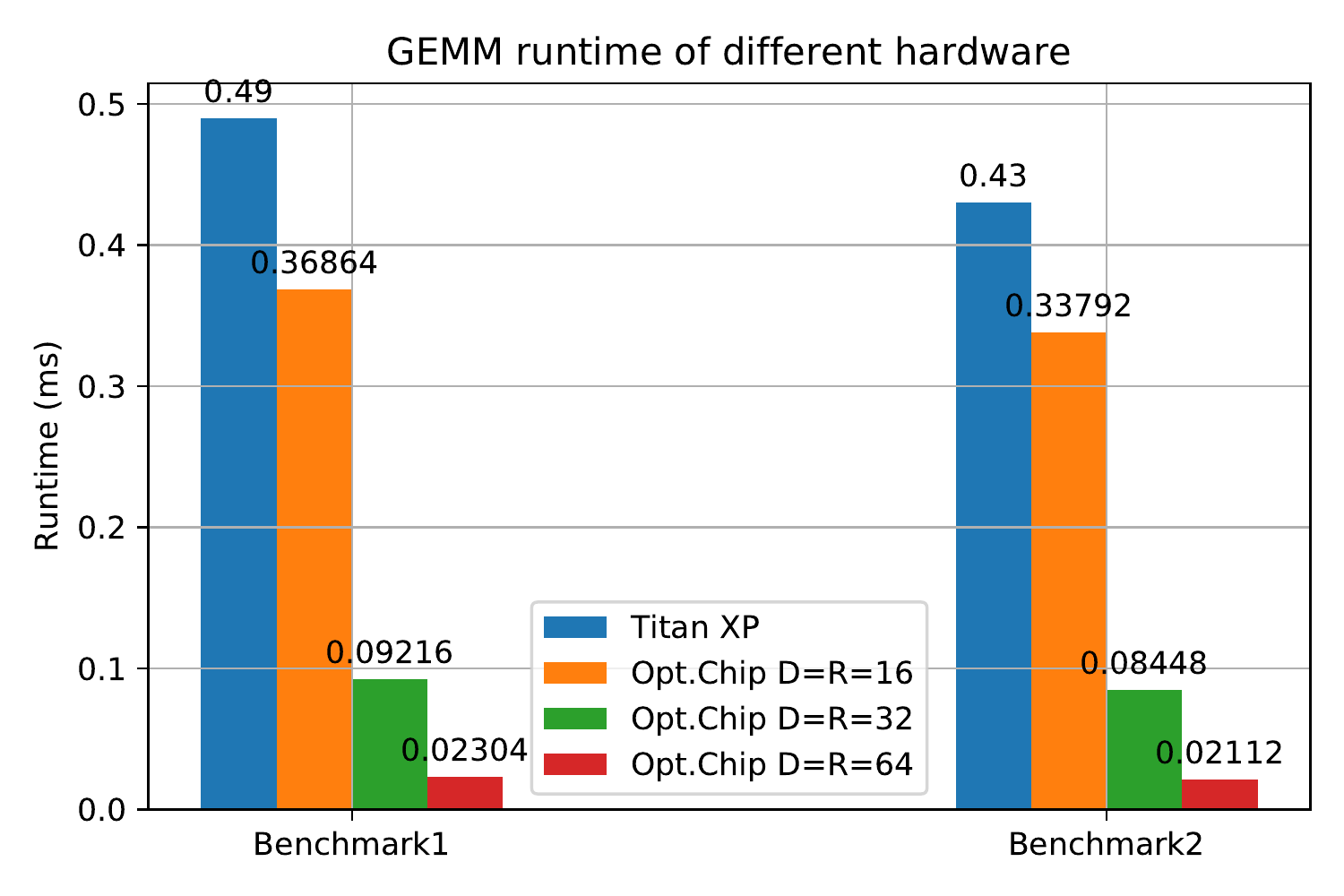}
  \caption{Runtime of the proposed photonic computing architecture as compared to Titan XP GPU \cite{deepbench}.}
\label{GEMM}
\end{figure}

In this work, General Matrix Multiplication (GEMM) is considered as a benchmark to evaluate the computation speed of the proposed architecture. Figure~\ref{GEMM} illustrates the computation time of the proposed photonic system as compared to the runtime of \textit{Titan XP} GPU \cite{deepbench}. The parameters of the benchmarked scenarios are the dimensions of the matrices as listed in Table~\ref{Table_Bench_Param}. As shown in Figure~\ref{GEMM}, by increasing the number of parallel waveguide channels and/or the number of MRMs in the proposed architecture, the processing time can be reduced notably. Moreover, Figure~\ref{GEMM} highlights the fact that the processing time of the photonic computing platform is lower than that of \textit{Titan XP} GPU, and the gap between the processing times exponentially increases when the photonic architecture scales up. 

\begin{table}
\begin{center}
\caption{Benchmarked parameters for  $\mathbf{A}_{m \times n} \times \mathbf{B}_{n \times k}$ .}
    \label{Table_Bench_Param}
 \begin{tabular}{||c c c c ||} 
 \hline
{Parameters} & {$m$} & {$n$}  & {$k$} \\ [0.5ex] 
 \hline
{Benchmark 1} & 7680 & 1500 & 2560 \\
 \hline
{Benchmark 2} & 10752 & 1 & 3584\\ [0.5ex] 
 \hline
\end{tabular}
\end{center}
\end{table}

\subsection{Massive-MIMO Wireless Communication Systems}
In this section, the performance and the efficiency of the proposed photonic computing system when employed in massive-MIMO wireless communication systems is studied. For this purpose, MMSE signal detection in an uplink scenario, where $K$ single-antenna users transmit their signals to a BS equipped with $M$ antennas, is considered. The Channel State Information (CSI) is assumed to be perfectly known at the transmitter and the receiver. The performance, in terms of the Symbol Error Rate (SER), and the processing time are compared with one of the most powerful GPUs, namely \textit{NVIDIA GeForce RTX 2080 Ti}\footnote{\url{https://www.nvidia.com/en-sg/geforce/graphics-cards/rtx-2080-ti/}}. 

\subsubsection{Time Efficiency Analysis}
\begin{figure}[htbp]
\centering
\includegraphics[width=0.9\textwidth]{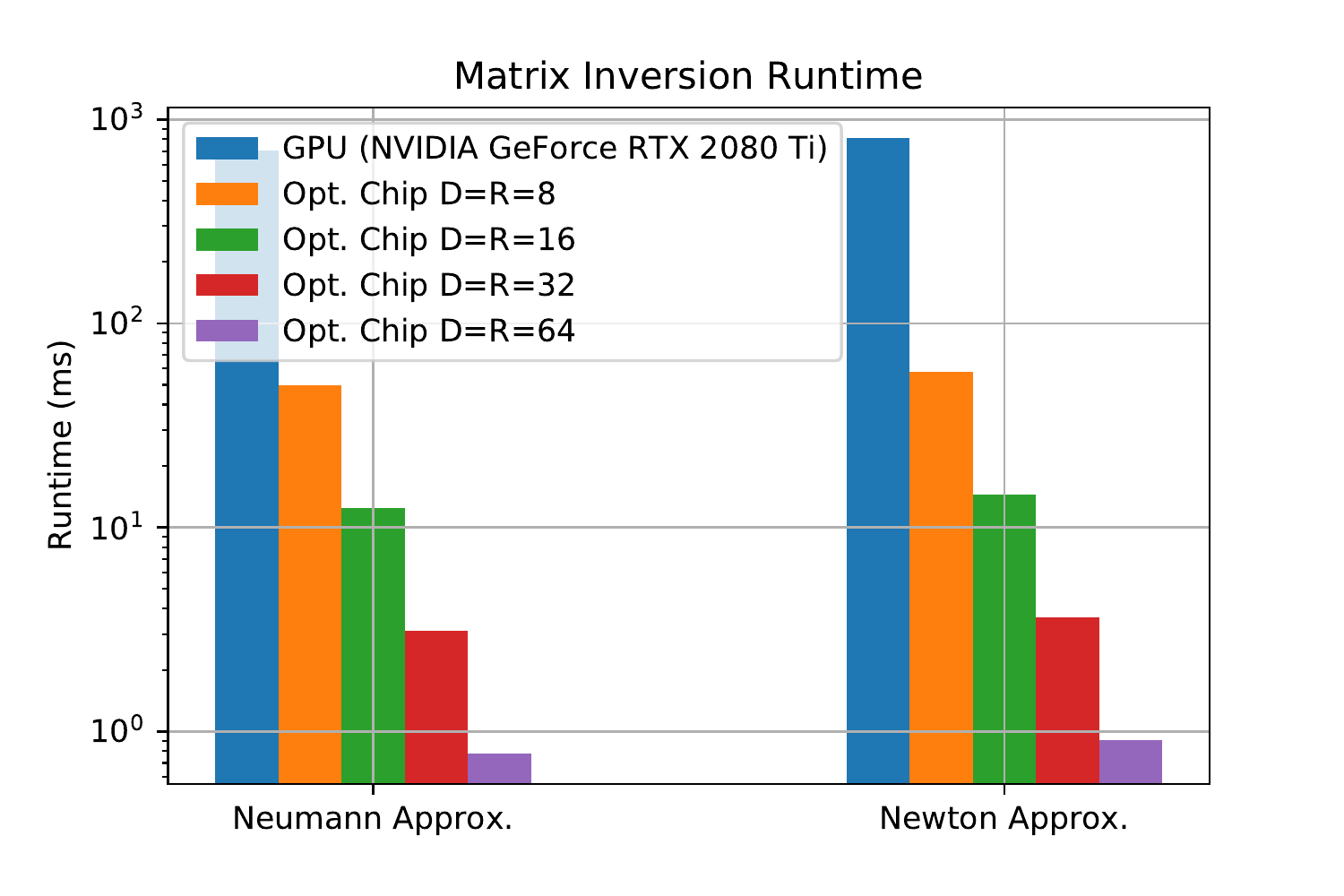}
 \caption{Matrix inversion processing time using Neumann series and Newton approximation approaches.}
\label{runtime}
\end{figure}

In this numerical study, a massive-MIMO system that consists of a 1024-antenna BS which is serving $K=64$ users is modelled. The processing time of this system for different parameters of the photonic system, i.e., different numbers of waveguide channels and different numbers of MRRs in each channel, is evaluated using \eqref{total_runtime}. Additionally, the computation time for two matrix inversion approximation methods, namely Neumann-series and Newton approximations, explained in Sections~\ref{Neumann} and \ref{Newton}, is reported.  As shown in Figure~\ref{runtime},  the processing time associated with the proposed photonic computing system is significantly less than that of GPU in both approximation approaches. Furthermore, as expected, increasing the number of parallel waveguide channels and modulators in each of those channels can further reduce the runtime. However, other limitations must be taken into account when increasing the number of components in the photonic system. It can be seen in Figure~\ref{runtime} that the computation time of Neumann-series approximation is marginally lower than that of the Newton approximation method. Moreover, the study in \cite{thanos2017algorithms} shows that both methods perform equally in terms of SER. Hence, Neumann approximation method is adopted in the next numerical experiments.

\subsubsection{Precision Analysis}
\begin{figure}[h]
\centering
\includegraphics[width=0.8\textwidth]{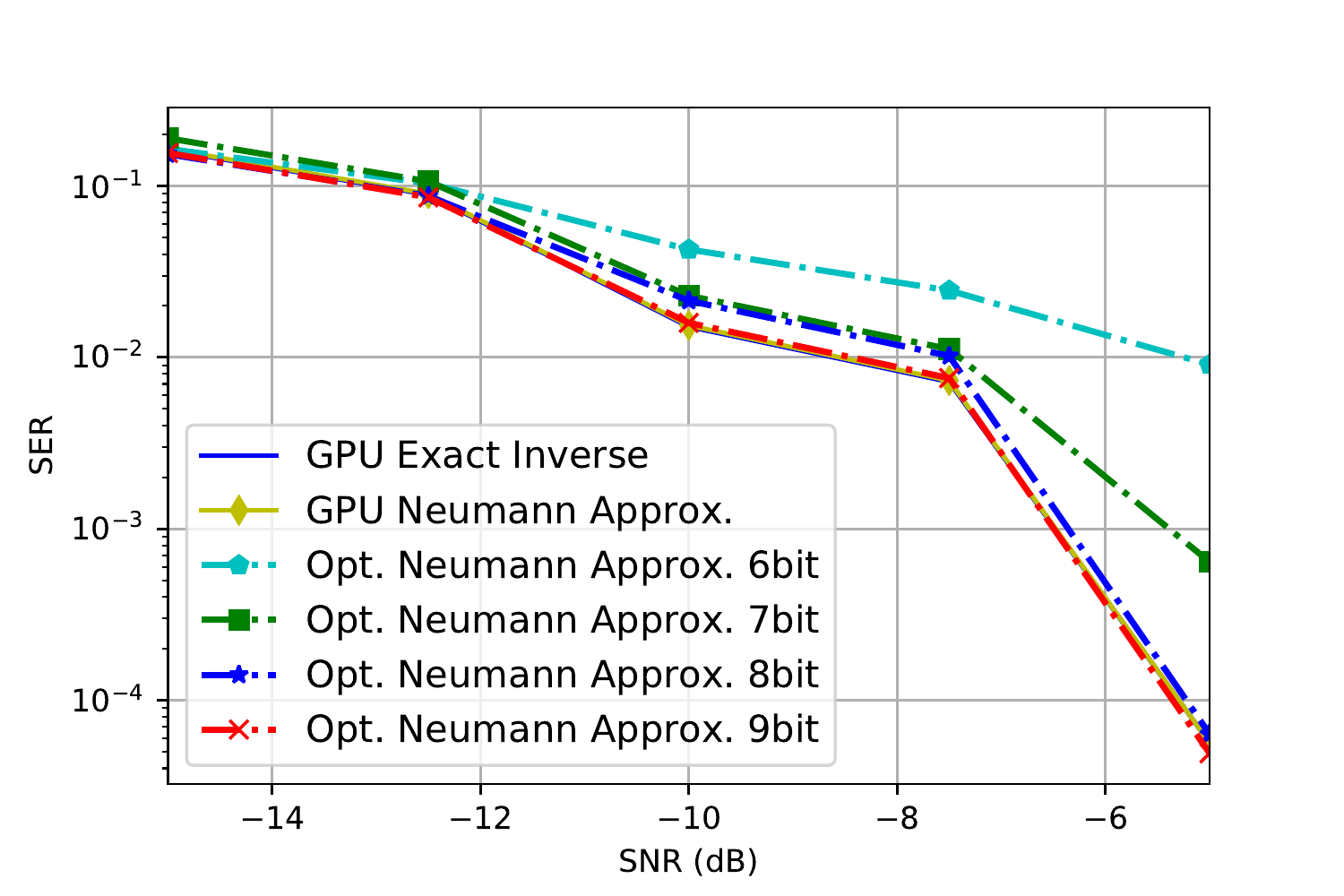}
 \caption{SER for MMSE detection in GPU and the proposed photonic computing platform for different precision bits. Neumann-series approximation is used for matrix inversion. }
\label{num_precision}
\end{figure}

In this section, we seek to evaluate the impact of the modulator control precision on the performance of the proposed optical platform. 
We consider a MIMO system with $K=8$ users being served with a BS equipped with $M=64$ antennas. The signal detection process in the BS is performed utilizing a photonic system with $D=8$ parallel waveguide channels, each of which with $2R=16$ MRRs. Figure~\ref{num_precision} illustrates the SER for different precisions over $10^5$ channel realizations. (In that figure GPU Exact indicates the performance of the GPU when the matrix inversion is computed using the built-in functions in the considered software, rather than approximating that using Neumann approximation method.) It can be seen that in the low-SNR regime, where the effect of transmission noise is dominant, the performance of the proposed photonic computing platform is similar to that of the GPU, regardless of the precision bits. However, as the SNR increases, the low-precision error becomes more dominant. As shown in Figure~\ref{num_precision}, for relatively higher SNR values, there is a notable gap between the performance of the photonic computing platform with only 6-bit modulator precision and that of the GPU. However, photonic computing can achieve the same performance as that of the GPU when the precision reaches to 8 bits (plus the sign bit). Therefore, in the next numerical experiments, an 8-bit precision is considered for the performance analysis. 

\subsubsection{Performance Analysis}
\begin{figure}[htbp]
\centering
\includegraphics[width=0.8\textwidth]{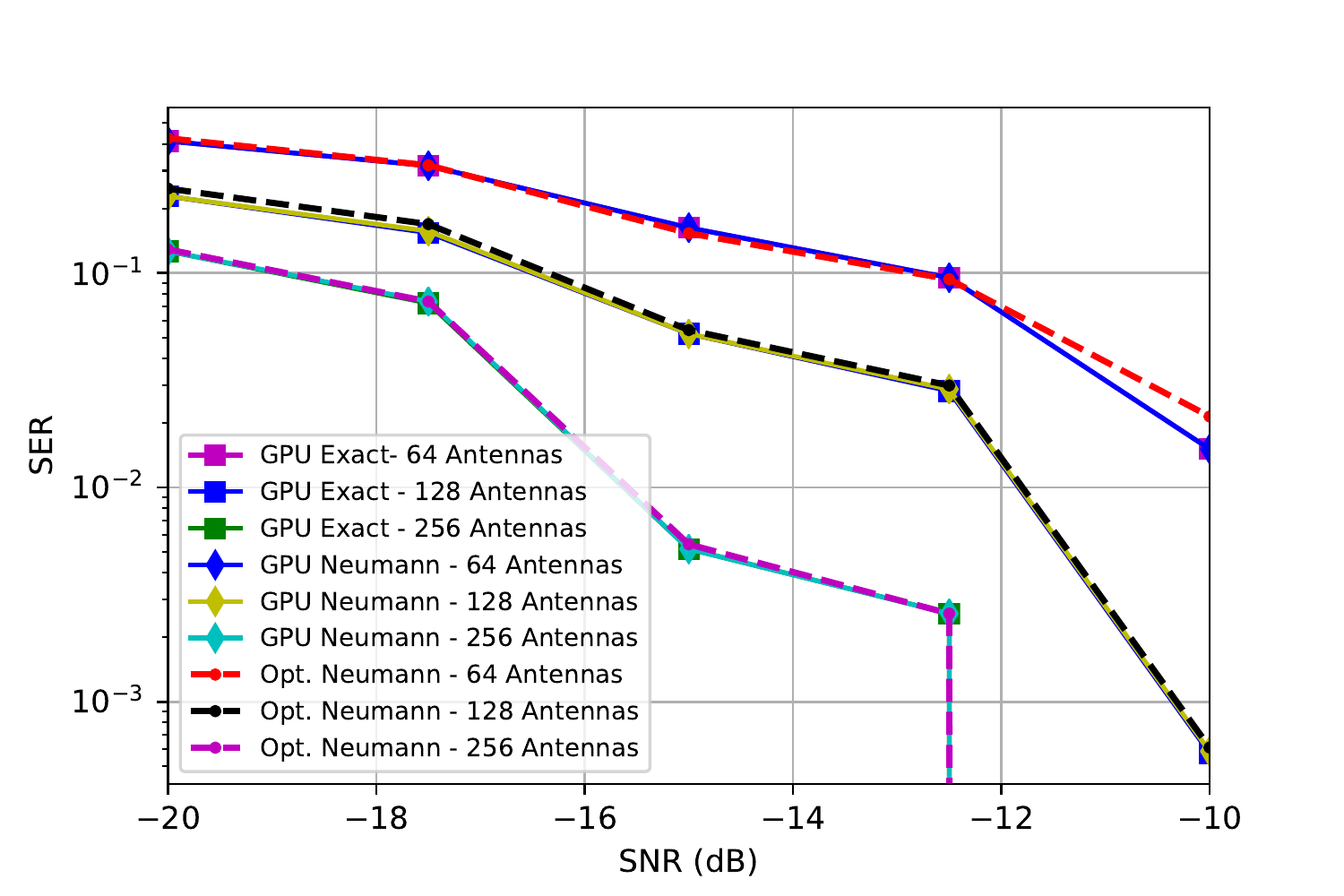}
 \caption{SER for MMSE detection in GPU and the proposed photonic computing platform for different number of BS antennas. Neumann-series approximation is used for matrix inversion.}
\label{Sweeping_Antenna}
\end{figure}

To evaluate the performance of the proposed photonic computing system when the number of BS antennas increases in a massive-MIMO wireless communication network, an uplink system with $K=8$ users and BS with different antenna array length is modelled. The signal detection in the BS is performed using MMSE detection method. Figure~\ref{Sweeping_Antenna} summarizes the simulation results for a photonic computing system with $D=8$ parallel waveguide channels, $2R=16$ MRRs in each channel, and 8-bit precision (plus the sign bit) over $10^5$ channel realizations. It can be seen in Figure~\ref{Sweeping_Antenna} that the proposed photonic computing platform has comparable performance to the GPU for all SNR values and different number of antennas. moreover, as expected, increasing the number of antenna elements (antenna array gain) can significantly improve the performance of the system in both cases. (Note that by increasing SNR, SER improves to the extent that for $\text{SNR}$ greater than -14~dB, no symbol error has been found.)  

Considering the above evaluation of the performance, power consumption, and the processing time of one of the most powerful GPUs, i.e., \textit{NVIDIA GeForce RTX 2080 Ti}, and those of the proposed optical computing platform, it can be seen that optical computing can provide the same performance in a relatively lower power consumption as conventional GPUs, while it can significantly reduce the processing time. That makes the proposed optical architecture as a promising candidate which is capable of supporting high computational complexity required in the next generations of wireless communication networks, while it can meet low latency and high reliability requirements of those systems.    

\section{Conclusion}
In this paper, a photonic computing architecture is proposed to be employed in next-generation massive-MIMO wireless communication systems to address their stringent computational requirements. The proposed computing approach is based on the B$\&$W architecture to implement MAC operations in the optical domain exploiting the light speed and lossless propagation. Preprocessing steps are developed so that the proposed computing system can represent matrices with any arbitrary values. Numerical experiments confirm that the proposed photonic computing architecture can offer the same performance as those of the most powerful digital electronics-based data processing units such as GPUs, while its time efficiency is shown to be several orders of magnitude better than that of the modern state-of-the-art GPUs. Based on the simulation results, the proposed photonic platform can be integrated into the 5GB base stations as a power- and cost-efficient solution to enable ultra-fast data processing for the next-generation wireless communication networks. 


\begin{backmatter}
\bmsection{Funding}
The research carried out for this work is self-funded.

\bmsection{Acknowledgments}
The  authors  would  like  to  thank  Dr. Bhavin J. Shastri,  assistant professor at Queen's University, Ontario, Canada,  for the fruitful discussions that the authors had with him. 

\bmsection{Disclosures}
The authors declare no conflicts of interest.

\end{backmatter}










\end{document}